\documentclass[12pt]{iopart}
\usepackage{iopams}  
\expandafter\let\csname equation*\endcsname\relax
\expandafter\let\csname endequation*\endcsname\relax
\usepackage{amsmath}

\usepackage{color}
\usepackage{graphicx}
\usepackage[T1]{fontenc}
\usepackage{units}
\usepackage{xspace}

\usepackage{cite}

\newcommand{\abs}[1]{\mathopen{}\mathclose\bgroup\left|#1\aftergroup\egroup\right|}
\newcommand{\norm}[1]{\mathopen{}\mathclose\bgroup\left\|#1\aftergroup\egroup\right\|}
\newcommand{\kl}[1]{\mathopen{}\mathclose\bgroup\left(#1\aftergroup\egroup\right)}
\newcommand{\klg}[1]{\mathopen{}\mathclose\bgroup\left\{#1\aftergroup\egroup\right\}}
\newcommand{\kle}[1]{\mathopen{}\mathclose\bgroup\left[#1\aftergroup\egroup\right]}
\newcommand{\kls}[1]{\mathopen{}\mathclose\bgroup\left\langle#1\aftergroup\egroup\right\rangle}

\renewcommand{\vec}[1]{\boldsymbol{#1}}

\newcommand{\del}[1]{\partial_{\obs_{#1}}}

\newcommand{\datapoints}{n\xspace}

\newcommand{\obs}{x}
\newcommand{\dobs}{\vec{\obs}\xspace}

\newcommand{\drift}{h}
\newcommand{\diffusion}{g}

\newcommand{\jumpsize}{s\xspace}
\newcommand{\jumpamplitude}{\xi\xspace}
\newcommand{\jumprate}{\lambda\xspace}

\newcommand{\umbrae}{\mathcal{U}\xspace}

\newcommand{\boundedrelativeerror}{\mathcal{R}_b\xspace}

\newcommand{\scalinga}{\alpha\xspace}
\newcommand{\scalingb}{\beta\xspace}
\newcommand{\scalingc}{\gamma\xspace}

\newcommand*\diff{\mathop{}\!\mathrm{d}\hspace{1pt}}
\newcommand{\dt}{\diff t}

\newcommand{\dx}{\diff \obs}
\newcommand{\dW}{\diff W}

\newcommand{\dJ}{\diff J}

\newcommand{\km}{KM\xspace}

\newcommand{\Ktwo}[2]{K^{(#1,#2)}}
\newcommand{\estKtwo}[2]{\hat{K}^{(#1,#2)}}

\newcommand{\Dtwo}[2]{D^{(#1,#2)}}

\newcommand{\atwo}[2]{A^{(#1,#2)}}
\newcommand{\btwo}[2]{B^{(#1,#2)}}
\newcommand{\ctwo}[2]{C^{(#1,#2)}}

\begin{document}

\title[Data-driven reconstruction of bivariate jump-diffusion models]{Enhancing the accuracy of a data-driven reconstruction of bivariate jump-diffusion models with corrections for higher orders of the sampling interval}

\author{Esra Aslim$^{1,2}$, Thorsten Rings$^{1,2}$, Lina Zabawa$^{1,2}$ and Klaus Lehnertz$^{1,2,3}$}
\address{$^1$ Department of Epileptology, University of Bonn Medical Centre, Venusberg Campus 1, 53127 Bonn, Germany}
\address{$^2$ Helmholtz Institute for Radiation and Nuclear Physics, University of Bonn, Nussallee 14--16, 53115 Bonn, Germany}
\address{$^3$ Interdisciplinary Center for Complex Systems, University of Bonn, Br{\"u}hler Stra\ss{}e 7, 53175 Bonn, Germany}
\ead{klaus.lehnertz@ukbonn.de}
\vspace{10pt}
\begin{indented}
\item[]\today
\end{indented}

%%%%%%%%%%%%%%%%%%%%%%%%%%%%%%%%%%%%%%%%%%%%%%%%%%%%%%%%%%%%%%%%%%%%%%%%
\begin{abstract} 
We evaluate the significance of a recently proposed bivariate jump-diffusion model for a data-driven characterization of interactions between complex dynamical systems.
For various coupled and non-coupled jump-diffusion processes, we find that the inevitably finite sampling interval of time-series data negatively affects the reconstruction accuracy of higher-order conditional moments that are required to reconstruct the underlying jump-diffusion equations.  
We derive correction terms for conditional moments in higher orders of the sampling interval and demonstrate their suitability to strongly enhance the data-driven reconstruction accuracy.
\end{abstract}
%%%%%%%%%%%%%%%%%%%%%%%%%%%%%%%%%%%%%%%%%%%%%%%%%%%%%%%%%%%%%%%%%%%%%%%%

\vspace{2pc}
\noindent{\it Keywords}: nonlinear dynamics, stochastic processes, interactions, higher-order corrections
%\maketitle

\frenchspacing

%%%%%%%%%%%%%%%%%%%%%%%%%%%%%%%%%%%%%%%%%%%%%%%%%%%%%%%%%%%%%%%%%%%%%%%%
\section{Introduction}
%%%%%%%%%%%%%%%%%%%%%%%%%%%%%%%%%%%%%%%%%%%%%%%%%%%%%%%%%%%%%%%%%%%%%%%%

The problem of reliably characterizing interactions between complex dynamical systems pervades many scientific fields.
Since real-world systems quite often impose restrictions to standard approaches, linear and nonlinear time-series-analysis techniques have been developed that allow one to estimate the strength, the direction, and the functional form of an interaction from pairs of time series of appropriate system observables.
Given that interactions can manifest themselves in various aspects of the dynamics, analysis techniques have been developed in diverse fields such as statistics, synchronization theory, nonlinear dynamics, information theory, and statistical physics (for an overview, see~\cite{pikovsky2001,kantz2003,reinsel2003,hlavackova2007,marwan2007,stankovski2017}).
Most of these techniques specifically concentrate on the (low-dimensional) deterministic part of the dynamics and have shown interactions to impact on amplitudes, phases, frequencies --~or even combinations thereof~-- as well as on trajectories in the respective phase spaces.
	
Many natural systems, however, exhibit both deterministic and stochastic features, even if the underlying dynamic is deterministic. 
Stochastic features in a system's dynamics may arise from a high number of degrees of freedom, from random forcing, and/or from nonlinear couplings. 
If the systems' dynamics can be described sufficiently by the Langevin equation, the first- and second-order Kramers-Moyal (KM) coefficients estimated from time-series data~\cite{friedrich2011,tabar2019book} can serve as an indicator for interactions between stochastic processes~\cite{prusseit2008a,lehle2013,scholz2017}.
A more general ansatz that accounts for the presence of 
additive as well as multiplicative diffusive fluctuations and discontinuous jump contributions in the time-series data~\cite{anvari2016,lehnertz2018,hashtroud2019} has been proposed only recently~\cite{gorjao2019}, namely a bivariate jump-diffusion model. 
It consists of two-dimensional diffusion and two-dimensional jumps that can be coupled to one another.
Such a model could improve theoretical modeling of time-series data e.g., in the neurosciences~\cite{ditlevsen2017,lombardi2020}, condensed matter physics~\cite{scalliet2015,plati2019,plati2020}, ecology~\cite{carpenter2011,li2015b}, or finance~\cite{ait2015}.

When analyzing empirical time-series data, one is faced with the issue of an inevitably finite sampling interval $\dt$, which not only influences the first- and second-order \km coefficients~\cite{gc5} but also causes non-vanishing higher-order ($>2$) ones.
For (one-dimensional) jump-diffusion processes, additional influences need to be taken into account~\cite{lehnertz2018}: jump events induce terms of order $\mathcal{O}(\dt)$ in the conditional moments
of even orders and the jump rate and amplitude induce terms of order $\mathcal{O}(\dt^2)$ in all conditional moments.
We here extend these studies and investigate the data-driven reconstruction of 
stochastic dynamical equations underlying interacting jump-diffusion processes with finite sampling interval.
We will show that in these cases, corrections for higher orders of the sampling interval strongly enhance reconstruction accuracy. 

The outline of this paper is as follows. 
In \sref{sec:methods}, we recall the definition of a bivariate jump-diffusion model, and we define our scale-independent measure to assess the accuracy of the reconstruction of conditional moments from time-series data. 
In \sref{sec:results}, we first illustrate the reconstruction of conditional moments of various jump-diffusion models with and without couplings from time-series data with finite sampling interval, thereby emphasizing the necessity for corrections for higher orders of the sampling interval. 
We then derive these corrections and demonstrate their suitability to enhance reconstruction accuracy. 
Finally, in \sref{sec:conclusions} we draw our conclusions.

%%%%%%%%%%%%%%%%%%%%%%%%%%%%%%%%%%%%%%%%%%%%%%%%%%%%%%%%%%%%%%%%%%%%%%%%
\section{Methods}
\label{sec:methods}
%%%%%%%%%%%%%%%%%%%%%%%%%%%%%%%%%%%%%%%%%%%%%%%%%%%%%%%%%%%%%%%%%%%%%%%%

\subsection{Bivariate jump-diffusion model}
\label{sec:2djdmod}

A bivariate jump-diffusion process consists of two-dimensional diffusion and two-dimensional jumps, that can be coupled to one another. 
It can be modeled via~\cite{anvari2016,tabar2019book,gorjao2019}
\begin{eqnarray}\label{eq:model} \fl
    \qquad\quad
		  \begin{pmatrix}\dx_1(t) \\ \dx_2(t)\end{pmatrix}
  & = \begin{pmatrix}\drift_1 \\ \drift_2\end{pmatrix}\dt
    + \begin{pmatrix}\diffusion_{1,1} & \diffusion_{1,2} \\ \diffusion_{2,1} & \diffusion_{2,2}\end{pmatrix}
		  \begin{pmatrix}\dW_1 \\ \dW_2\end{pmatrix}
  & + \begin{pmatrix}\jumpamplitude_{1,1} & \jumpamplitude_{1,2} \\ \jumpamplitude_{2,1} & \jumpamplitude_{2,2}\end{pmatrix}
	    \begin{pmatrix}\dJ_1 \\ \dJ_2\end{pmatrix}.
\end{eqnarray}
The drift is a two-dimensional vector $\vec{\drift} = (\drift_1, \drift_2)$ with $\vec{\drift} \in \mathbb{R}^2$, where each dimension of $\vec{\drift}$, i.e., $\drift_i$, may depend on state variables $\obs_1(t)$ and $\obs_2(t)$.
The diffusion takes a matrix ${\vec{\diffusion} \in \mathbb{R}^{2\times 2}}$ with the diagonal elements of $\vec{\diffusion}$ comprise the diffusion coefficients of self-contained stochastic diffusive processes. 
The off-diagonal elements represent interdependencies between the two Wiener processes $\vec{W}=(W_1,W_2)$, i.e., they result from an interaction between the two processes.
The Wiener processes act as independent Brownian noises for the state variables with $\kls{\dW_i} = 0, \kls{\dW_i^2} = \dt, \forall i$.
The discontinuous jump terms are contained in $\vec{\xi} \in \mathbb{R}^{2\times 2}$ and $\vec{\dJ} \in \mathbb{N}^{2}$, where $\vec{\dJ}$ represents a two-dimensional Poisson process.
These are Poisson-distributed jumps with an average jump rate $\vec{\jumprate} \in \mathbb{R}^2$ in unit time $t$.
The average expected number of jumps of each jump process  $J_i$ in a timespan $t$ is $\jumprate_i t$.
The jump amplitudes $\vec{\jumpamplitude}$ are Gaussian distributed with zero mean and standard deviation (or size) $\jumpsize_{i,j}$.
We note that elements of vectors $\vec{\drift}$ and $\vec{\dJ}$ as well as of matrices $\vec{\diffusion}$ and $\vec{\jumpamplitude}$ may, in general, be state- and time-dependent; for convenience of notation, we omit these dependencies.

The two-dimensional \km coefficients of orders $(\ell, m)$ of a bivariate process read 
\begin{eqnarray}
\Dtwo{\ell}{m}(\dobs,t,\dt)&=\frac{1}{(\ell+m)!}\lim_{\dt\to 0}\!\frac{1}{\dt}\Ktwo{\ell}{m}(\dobs,t,\dt)
\end{eqnarray}
with the conditional moments
\begin{eqnarray}
\Ktwo{\ell}{m}(\dobs,t,\dt)=\kls{[\obs_1(t+\dt)-\obs_1(t)]^{\ell}[\obs_2(t+\dt)-\obs_2(t)]^{m}}\Big|_{\substack{\obs_1(t)=x_1 \\ \obs_2(t)=x_2}}
\label{eq:conditionalmoments_definition}
\end{eqnarray}
which can be directly estimated from time-series data~\cite{gorjao2019}. 
They are related to the elements of the drift vector, the elements of the diffusion matrix, and the jump components via:
\begin{eqnarray}\label{eq:KMtheor}
\Ktwo{1}{0}(\dobs,t,\dt) &= \drift_1 \dt \\  \nonumber
\Ktwo{0}{1}(\dobs,t,\dt) &= \drift_2 \dt \\ \nonumber
\Ktwo{1}{1}(\dobs,t,\dt) &= \Big[\diffusion_{11}\diffusion_{21}+\diffusion_{12}\diffusion_{22}\Big] \dt \\ \nonumber
\Ktwo{2}{0}(\dobs,t,\dt) &= \Big[\diffusion_{11}^2+\jumpsize_{11}\jumprate_1+\diffusion_{12}^2+\jumpsize_{12}\jumprate_2\Big] \dt \\ \nonumber
\Ktwo{0}{2}(\dobs,t,\dt) &= \Big[\diffusion_{21}^2+\jumpsize_{21}\jumprate_1+\diffusion_{22}^2+\jumpsize_{22}\jumprate_2\Big] \dt \\ \nonumber
\Ktwo{2\ell}{2m}(\dobs,t,\dt) &= \Big[\jumpsize_{11}^{\ell}\jumpsize_{21}^{m}\jumprate_1+\jumpsize_{12}^{\ell}\jumpsize_{22}^{m}\jumprate_2\Big]\frac{(2\ell)!}{2^{\ell} \ell!}\frac{(2m)!}{2^{m}m!} \dt ,
\end{eqnarray}
with $(\ell,m) \in \mathbb{N}^+$. 
We omitted the state- and time-dependencies in the drift and diffusion functions and jump components to enhance readability.
If such processes are coupled, the elements of the drift vector, the elements of the diffusion matrix, and the jump components may depend on both state variables and thus information about the coupling may be contained in the respective conditional moments.
However, apart from the drift functions, there is --~to our knowledge~-- not yet an analytical way to estimate all diffusion functions and jump components in order to characterize interactions in the diffusion and jump part of the dynamics. 
As an alternative, one can use some optimization techniques to estimate these unknown functions~\cite{tabar2019book}.

\subsection{Assessing the accuracy of a data-driven reconstruction of conditional moments}
\label{sec:errorestimators}

Following Refs.~\cite{prusseit2008a,gorjao2019}, we employ a \textit{distance} measure to relate theoretical and numerical results and to quantify the deviation of the obtained conditional moments from the functions employed.
In order to allow a comparison of the accuracy of the reconstruction of conditional moments of order $(\ell,m)$, we here use a scale-independent distance measure, $\umbrae^{(\ell,m)}$, which is based on the bounded relative error of the difference between estimated and theoretical conditional moments (see~\ref{app:measures} for details).
$\umbrae^{(\ell,m)}<1$ indicates a sufficient accuracy of the reconstruction of the conditional moment of order $(\ell,m)$.

We estimate conditional moments up to orders $\ell = m = 6$ from normalized time series (zero mean and unit variance) that we obtain from numerically integrating bivariate jump-diffusion equations (Euler-Maruyama scheme~\cite{kloeden1999} with a sampling interval $\dt = 10^{-3}$).
Time series consist of $\datapoints=10^7$ data points (after eliminating $5 \times 10^3$ transients), and we ensure that individual jump numbers $n_j \simeq \jumprate\datapoints$ varied by at most 10\,\% for a constant jump rate.

%%%%%%%%%%%%%%%%%%%%%%%%%%%%%%%%%%%%%%%%%%%%%%%%%%%%%%%%%%%%%%%%%%%%%%%%
\section{Results}
\label{sec:results}
%%%%%%%%%%%%%%%%%%%%%%%%%%%%%%%%%%%%%%%%%%%%%%%%%%%%%%%%%%%%%%%%%%%%%%%%

For our investigations, we consider various interacting jump-diffusion processes. 
We begin with reconstructing conditional moments of a bivariate jump-diffusion model with uni-directional couplings in the drift and in the diffusion from time-series data.
Since interactions between jump-diffusion processes may lead to over- and underrepresented parts of the dynamics~\cite{gorjao2019}, we then reconstruct conditional moments of a bivariate jump-diffusion model with a disproportionally weighted drift, diffusion, and jump part from time-series data.
\begin{figure}
\centering
\includegraphics{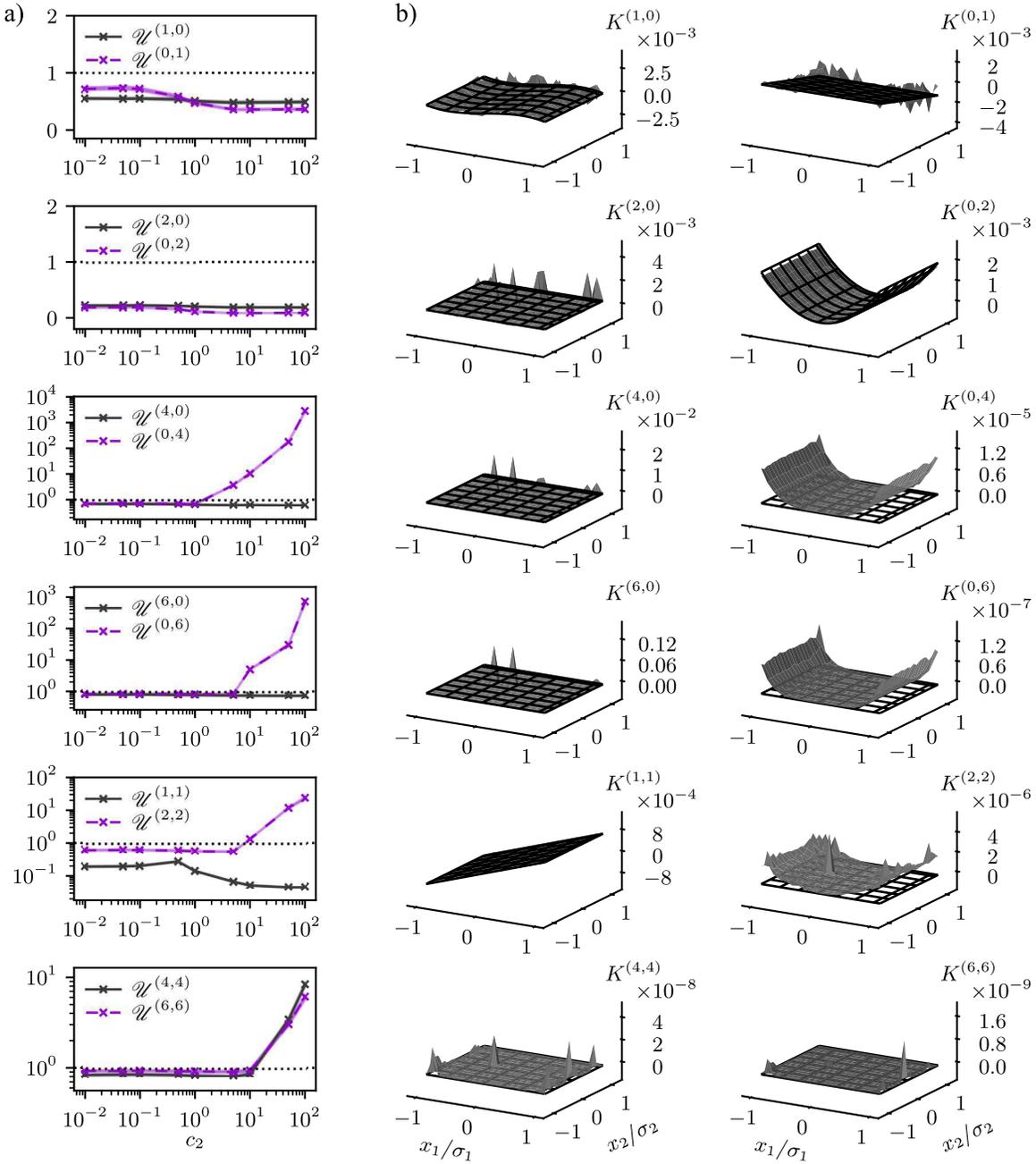}
\caption{
a) Reconstruction accuracy ($\umbrae^{(\ell,m)}$) for conditional moments up to order $\ell=m=6$ of a bivariate jump-diffusion model with a uni-directional coupling in the diffusion part (see equation~\eref{eq:2djdcoupled}) for various values of coupling strength $c_2$.
$\umbrae^{(\ell,m)}< 1$ (horizontal dotted line) indicates a sufficient accuracy.
Medians and interquartile ranges (shaded area) derived from 50 time series generated with the respective models using random initial conditions.
Lines are for eye guidance only.
b) Two-dimensional theoretical conditional moments (equation~\eref{eq:KMtheor}) up to order $\ell=m=6$ (black grid) and reconstructed ones (gray surface) from exemplary time series generated using equation~\eref{eq:2djdcoupled} with $c_1=0$ and $c_2=100$.
Reconstruction accuracy is insufficient for conditional moments of orders $(2,2)$, $(0,4)$, $(4,4)$, $(0,6)$, and $(6,6)$ ($\umbrae^{(2,2)}\approx23$, $\umbrae^{(0,4)}\approx2\,639$, $\umbrae^{(4,4)}\approx9$, $\umbrae^{(0,6)}\approx682$, and $\umbrae^{(6,6)}\approx7$).
}
\label{fig:example_coupling}
\end{figure}

%%%%%%%%%%%%%%%%%%%%%%%%%%%%%%%%%%%%%%%%%%%%%%%%%%%%%%%%%%%%%%%%%%%%%%%%
\subsection{Jump-diffusion model with uni-directional couplings}
\label{sec:unidir-couplings}
%%%%%%%%%%%%%%%%%%%%%%%%%%%%%%%%%%%%%%%%%%%%%%%%%%%%%%%%%%%%%%%%%%%%%%%%

We examine uni-directional couplings in the drift and in the diffusion of an exemplary bivariate jump-diffusion process modeled via:
\begin{eqnarray}\label{eq:2djdcoupled} \fl
		  \begin{pmatrix}\dx_1(t) \\ \dx_2(t) \end{pmatrix}
  & = &\begin{pmatrix}-x_1^3+x_1 \\ -x_2 + f_1(x_1)\end{pmatrix}\dt
    + \begin{pmatrix}0.1 & 0.5 \\ 0.3 & 0.2 + f_2(x_1)\end{pmatrix}
		  \begin{pmatrix}\dW_1 \\ \dW_2\end{pmatrix} \\ \nonumber
  & + &\begin{pmatrix}\jumpamplitude_{1,1} & \jumpamplitude_{1,2} \\ \jumpamplitude_{2,1} & \jumpamplitude_{2,2}\end{pmatrix}
	    \begin{pmatrix}\dJ_1 \\ \dJ_2\end{pmatrix}, 
\end{eqnarray}
with
\begin{eqnarray*}
\vec{\jumpsize} = \begin{pmatrix}0.2 & 0.5 \\ 0.3 & 0.1\end{pmatrix}, \,\, \vec{\jumprate} = \begin{pmatrix}0.1 \\ 0.3\end{pmatrix} 
\end{eqnarray*}
and with the coupling terms $f_1(x_1)=c_1x_1$ and $f_2(x_1)=c_2x_1$. 
We vary the coupling strengths $c_1$ and $c_2$ each over four orders of magnitude and reconstruct conditional moments from 50 time series with random initial conditions and with finite sampling interval.
For couplings in the drift part ($c_1\in[0.01,100]$; $c_2=0$), almost all conditional moments up to order $\ell=m=6$ can be reconstructed with sufficient accuracy (data not shown). 
An exception builds the conditional moment $\Ktwo{0}{2}$ with diffusion contributions of process $\obs_2$. 
For this moment, we observe an insufficient reconstruction accuracy ($\umbrae^{(0,2)}>1$) in case of strong couplings ($c_1>50$).
For couplings in the diffusion part ($c_2\in[0.01,100]$; $c_1=0$; see~\fref{fig:example_coupling}), we observe at large values of the coupling strength ($c_2>1$) rather strong inaccuracies for some conditional moments.
These include conditional moments with jump contributions of process $\obs_2$, namely $\Ktwo{0}{4}$, $\Ktwo{0}{6}$, and $\Ktwo{i}{i}$ with $i\in\{2,4,6\}$ (\fref{fig:example_coupling}a). 
A visual inspection of these conditional moments for a coupling in the diffusion part with $c_2=100$ (\fref{fig:example_coupling}b) indicates that the theoretical moments  (equation~\eref{eq:KMtheor}) fail to account for a dependency on process~$\obs_1$.
This dependency on process~$\obs_1$ appears similar to the dependency of $\Ktwo{0}{2}$ on $\obs_1$, where we note that $\Ktwo{0}{2}$ contains information about the coupling in the diffusion part (see equation~\eref{eq:KMtheor}).

%%%%%%%%%%%%%%%%%%%%%%%%%%%%%%%%%%%%%%%%%%%%%%%%%%%%%%%%%%%%%%%%%%%%%%%%
\subsection{Jump-diffusion model with disproportionally weighted parts}
\label{sec:dispropweights}
%%%%%%%%%%%%%%%%%%%%%%%%%%%%%%%%%%%%%%%%%%%%%%%%%%%%%%%%%%%%%%%%%%%%%%%%
%
\begin{figure}[b]
\centering
\includegraphics{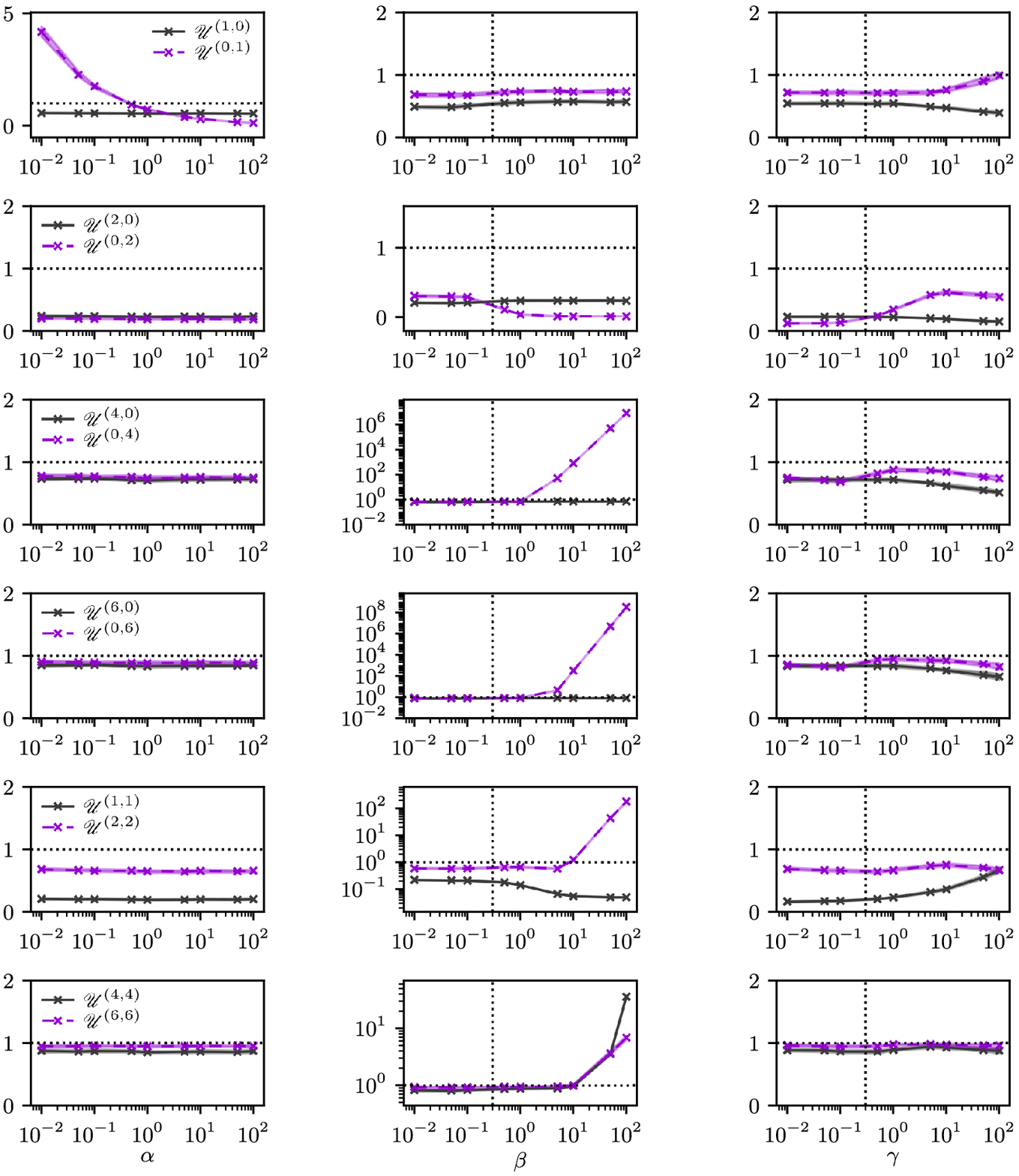}
\caption{
Reconstruction accuracy ($\umbrae^{(\ell,m)}$) of conditional moments up to order $\ell=m=6$ of a bivariate jump-diffusion model with disproportionally weighted parts (see equation~\eref{eq:2djdmixing})
for various values of drift-, diffusion-, and jump-scaling parameters ($\scalinga$, $\scalingb$, $\scalingc$). 
The vertical dotted line at $\scalingb=g_{2,1}=0.3$ and $\scalingc=0.3$ indicates the point where the diffusion-scaling parameter is equal to the jump-scaling parameter.
The horizontal dotted line indicates a sufficient accuracy ($\umbrae^{(\ell,m)}< 1$). 
Medians and interquartile ranges (shaded area) derived from 50 time series generated with the respective models using random initial conditions.
Lines are for eye guidance only.
}
\label{fig:results_umbrae}
\end{figure}
Next, we examine processes derived from an exemplary bivariate jump-diffusion model with disproportionally weighted parts, which we obtain by rescaling drift, diffusion, and jump dynamics. 
For the latter two, we allow for a mixing of the Wiener processes and a mixing of the Poisson processes (non-vanishing off-diagonal elements in diffusion and jump size matrix):
\begin{eqnarray}\label{eq:2djdmixing} \fl
    \qquad\quad
		  \begin{pmatrix}\dx_1(t) \\ \dx_2(t) \end{pmatrix}
  & = \begin{pmatrix}-x_1^3+x_1 \\ -\scalinga x_2\end{pmatrix}\dt
    + \begin{pmatrix}0.1 & 0.5 \\ \scalingb & 0.2\end{pmatrix}
		  \begin{pmatrix}\dW_1 \\ \dW_2\end{pmatrix}
  & + \begin{pmatrix}\jumpamplitude_{1,1} & \jumpamplitude_{1,2} \\ \jumpamplitude_{2,1} & \jumpamplitude_{2,2}\end{pmatrix}
	    \begin{pmatrix}\dJ_1 \\ \dJ_2\end{pmatrix}, 
\end{eqnarray}
with
\begin{eqnarray*}
\vec{\jumpsize} = \begin{pmatrix}0.2 & 0.5 \\ \scalingc & 0.1\end{pmatrix}, \,\, \vec{\jumprate} = \begin{pmatrix}0.1 \\ 0.3\end{pmatrix}. 
\end{eqnarray*}
Here $\scalinga$ denotes the drift-, $\scalingb$ the diffusion- and $\scalingc$ the jump-scaling parameter.
We rescale the jump size since it equals the variance of the Gaussian-distributed jump amplitudes and thus weighs the jump part of the dynamics.
We vary each scaling parameter over four orders of magnitude (while keeping the others fixed) and reconstruct conditional moments of the model from 50 time series with random initial conditions and with finite sampling interval.

When rescaling the drift part (with $\scalingb = 0.3$ and $\scalingc = 0.3$), almost all conditional moments up to order $\ell=m=6$ can be reconstructed with sufficient accuracy (\fref{fig:results_umbrae}). 
The inaccuracy seen for $\Ktwo{0}{1}$ for $\scalinga < 1$ is to be expected given our chosen bivariate jump-diffusion model.
If we rescale the diffusion part (with $\scalinga = 1$ and $\scalingc = 0.3$), we observe at large values of the scaling parameter ($\scalingb >1$) rather strong inaccuracies for conditional moments $\Ktwo{0}{4}$, $\Ktwo{0}{6}$, and $\Ktwo{i}{i}$ with $i\in\{2,4,6\}$ (\fref{fig:results_umbrae}).
As with jump-diffusion models with uni-directional couplings (\sref{sec:unidir-couplings}), these conditional moments contain jump contributions of process $\obs_2$.
Rescaling the jump part (with $\scalinga = 1$ and $\scalingb = 0.3$) has no effect for the considered range of values here ($\scalingc\in[0.01,100]$; see \fref{fig:results_umbrae}).

%%%%%%%%%%%%%%%%%%%%%%%%%%%%%%%%%%%%%%%%%%%%%%%%%%%%%%%%%%%%%%%%%%%%%%%%
\subsection{Corrections for higher orders of the sampling interval}
\label{sec:correction}
%%%%%%%%%%%%%%%%%%%%%%%%%%%%%%%%%%%%%%%%%%%%%%%%%%%%%%%%%%%%%%%%%%%%%%%%
Our findings presented above demonstrate that uni-directional couplings in the diffusion may have a similar impact on the accuracy of the reconstruction of conditional moments as rescaling the diffusion part.
Estimated conditional moments differ from the respective theoretical moments with jump contributions of process $\obs_2$  at large values of either the coupling strength or the scaling parameter.
Given these observations and since we know that a finite sampling interval may have a non-negligible impact on the reconstruction of higher-order conditional moments of one-dimensional jump-diffusion models~\cite{lehnertz2018} from time-series data, we conjecture that a similar impact can be expected for bivariate jump-diffusion models.

To test this conjecture, we derive the theoretical conditional moments for different orders of the sampling interval $\dt$ of a bivariate jump-diffusion model (the derivation and further expressions of conditional moments can be found in \ref{app:timeintervalderivation-jumps}).
With the abbreviations
\begin{eqnarray*}
\atwo{1}{0} &= h_1\\
\atwo{0}{1} &= h_2\\
\btwo{1}{1} &= \frac{1}{2}\Big[\diffusion_{11}\diffusion_{21}+\diffusion_{12}\diffusion_{22}\Big] \\
\btwo{2}{0} &= \frac{1}{2}\Big[\diffusion_{11}^2+\jumpsize_{11}\jumprate_1+\diffusion_{12}^2+\jumpsize_{12}\jumprate_2\Big] \\
\btwo{0}{2} &= \frac{1}{2}\Big[\diffusion_{21}^2+\jumpsize_{21}\jumprate_1+\diffusion_{22}^2+\jumpsize_{22}\jumprate_2\Big] \\
\ctwo{2\ell}{2m} &=  \frac{1}{(2\ell+2m)!}\Big[\jumpsize_{11}^{\ell}\jumpsize_{21}^{m}\jumprate_1+\jumpsize_{12}^{\ell}\jumpsize_{22}^{m}\jumprate_2\Big]\frac{(2\ell)!}{2^{\ell} \ell!}\frac{(2m)!}{2^{m}m!},
\end{eqnarray*}
where $(\ell,m) \in \mathbb{N}^+$,
the theoretical conditional moments of orders $(0,4)$ and $(2,2)$ with correction terms up to order $\mathcal{O}(\dt^2)$ read
\begin{eqnarray*}
\fl\Ktwo{0}{4}(\dobs,t,\dt) = & 4!\ctwo{0}{4}\dt\\
& + \frac{1}{2}\Big[  4!\big(\btwo{0}{2}\big)^2
+ 4!\big(\atwo{0}{1}\del{2}\ctwo{0}{4} + \atwo{1}{0}\del{1}\ctwo{0}{4}\big)\\
& + 4\cdot4!\ctwo{0}{4}\del{2}\atwo{0}{1}
+ 4!\big( \btwo{0}{2}\del{2}^2\ctwo{0}{4} + \btwo{2}{0}\del{1}^2\ctwo{0}{4}\\
& + 2\btwo{1}{1}\del{1}\del{2}\ctwo{0}{4} \big)
+ 6\cdot4!\big( \ctwo{0}{4}\del{2}^2\btwo{0}{2} + \ctwo{2}{2}\del{1}^2\btwo{0}{2} \big)\\
& + 4\cdot5! \big( \ctwo{0}{6}\del{2}^3\atwo{0}{1} + 3\ctwo{2}{4}\del{1}^2\del{2}\atwo{0}{1} \big)\\
& + 6\cdot4!\ctwo{2}{2}\del{1}^2\del{2}^2\ctwo{0}{4}
 + 3\cdot6!\ctwo{2}{4}\del{1}^2\del{2}^2\btwo{0}{2}
+ \mathcal{O}(\delta)\Big]\dt^2 + \mathcal{O}(\dt^3)
\end{eqnarray*}
and
\begin{eqnarray*}
\fl\Ktwo{2}{2}(\dobs,t,\dt) = & 4!\ctwo{2}{2}\dt\\
& + \frac{1}{2}\Big[ \frac{4!}{3}\big(\btwo{2}{0}\btwo{0}{2} + 2\big(\btwo{1}{1}\big)^2\big)\\
& + 4!\big(\atwo{1}{0}\del{1}\ctwo{2}{2} + \atwo{0}{1}\del{2}\ctwo{2}{2}\big)\\
& + 2\cdot4!\big( \ctwo{2}{2}\del{1}\atwo{1}{0} + \ctwo{2}{2}\del{2}\atwo{0}{1} \big)\\
& + 4!\big( \btwo{2}{0}\del{1}^2\ctwo{2}{2} + \btwo{0}{2}\del{2}^2\ctwo{2}{2} + 2\btwo{1}{1}\del{1}\del{2}\ctwo{2}{2} \big)\\
& + 4!\big( \ctwo{4}{0}\del{1}^2\btwo{0}{2} + \ctwo{2}{2}\del{2}^2\btwo{0}{2} + \ctwo{2}{2}\del{1}^2\btwo{2}{0}  + \ctwo{0}{4}\del{2}^2\btwo{2}{0}\\
& + 8\ctwo{2}{2}\del{1}\del{2}\btwo{1}{1} \big)
+ 2\cdot\frac{6!}{3!}\big( \ctwo{4}{2}\del{1}^3\atwo{1}{0} + 3\ctwo{2}{4}\del{1}\del{2}^2\atwo{1}{0}\\
& + \ctwo{2}{4}\del{2}^3\atwo{0}{1}
+ 3\ctwo{4}{2}\del{1}^2\del{2}\atwo{0}{1} \big)
+ 6\cdot4!\ctwo{2}{2}\del{1}^2\del{2}^2\ctwo{2}{2}\\
& + \frac{5!}{2!}\big( 6\big(\ctwo{4}{2}\del{1}^2\del{2}^2\btwo{0}{2} + \ctwo{2}{4}\del{1}^2\del{2}^2\btwo{2}{0}\big) \\
& + 16\big(\ctwo{4}{2}\del{1}^3\del{2}\btwo{1}{1} + \ctwo{2}{4}\del{1}\del{2}^3\btwo{1}{1}\big) \big)
+ \mathcal{O}(\delta)\Big]\dt^2 + \mathcal{O}(\dt^3).
\end{eqnarray*}
For the differential operator, we use the short notation $\del{i}=\frac{\partial}{\partial\obs_i}$.
With $\mathcal{O}(\delta)$, we indicate all terms that contain $\ctwo{\ell}{m}$ of higher-order or derivatives $\del{i}^j$, $j>3$,
and with $\mathcal{O}(\dt^3)$ all terms that contain higher orders $(\ge3)$ of the sampling interval $\dt$.
We note that the terms of order $\mathcal{O}(\dt^2)$ can introduce drift, diffusion and jump contributions to each conditional moment. 
\begin{figure}[b]
\includegraphics{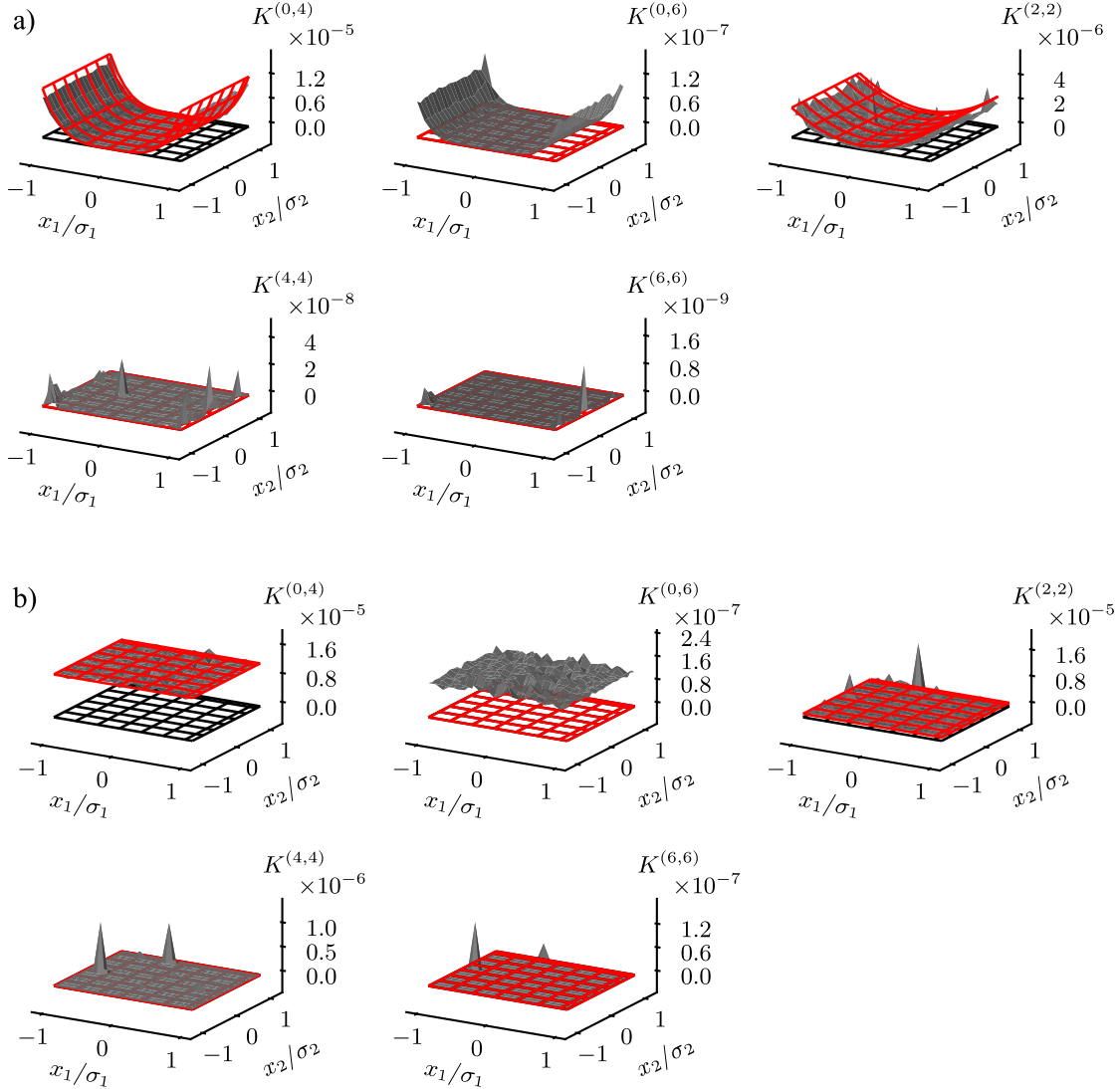}
\caption{a) Same as~\fref{fig:example_coupling}b) but in addition we present theoretical conditional moments, for which we considered terms of $\mathcal{O}(\dt^2)$ (red grid).
In this case, we obtain a sufficient quality of the reconstruction of conditional moments of all shown orders except of $(0,6)$, $(4,4)$ and $(6,6)$ ($\umbrae_\text{corr}^{(0,6)}\approx37$, $\umbrae_\text{corr}^{(4,4)}\approx9$ and $\umbrae_\text{corr}^{(6,6)}\approx7$). 
b) Same as~a) but for a bivariate jump-diffusion model with a more weighted diffusion part (see equation~\eref{eq:2djdmixing} with $\scalinga=1$, $\scalingb=100$ and $\scalingc=0.3$).
By considering terms of order $\mathcal{O}(\dt^2)$ in the theoretical conditional moments, we obtain a sufficient quality of the reconstruction of conditional moments of all orders except of $(0,6)$, $(4,4)$ and $(6,6)$ ($\umbrae_\text{corr}^{(0,6)}\approx2\,736\,249$, $\umbrae_\text{corr}^{(4,4)}\approx41$, $\umbrae_\text{corr}^{(6,6)}\approx8$).
}
\label{fig:example_corrected}
\end{figure}

With these correction terms, we now focus on conditional moments, which are affected by a uni-directional coupling in the diffusion or a more weighted diffusion part. 
Already a visual inspection reveals that considering terms of order~$\mathcal{O}(\dt^2)$ can clearly improve the reconstruction of some conditional moments (see \fref{fig:example_corrected}; $\umbrae_{\text{corr}}^{(\ell,m)}$ indicates the distance measure for which correction terms were considered).
\begin{figure}[ht]
\centering
\includegraphics{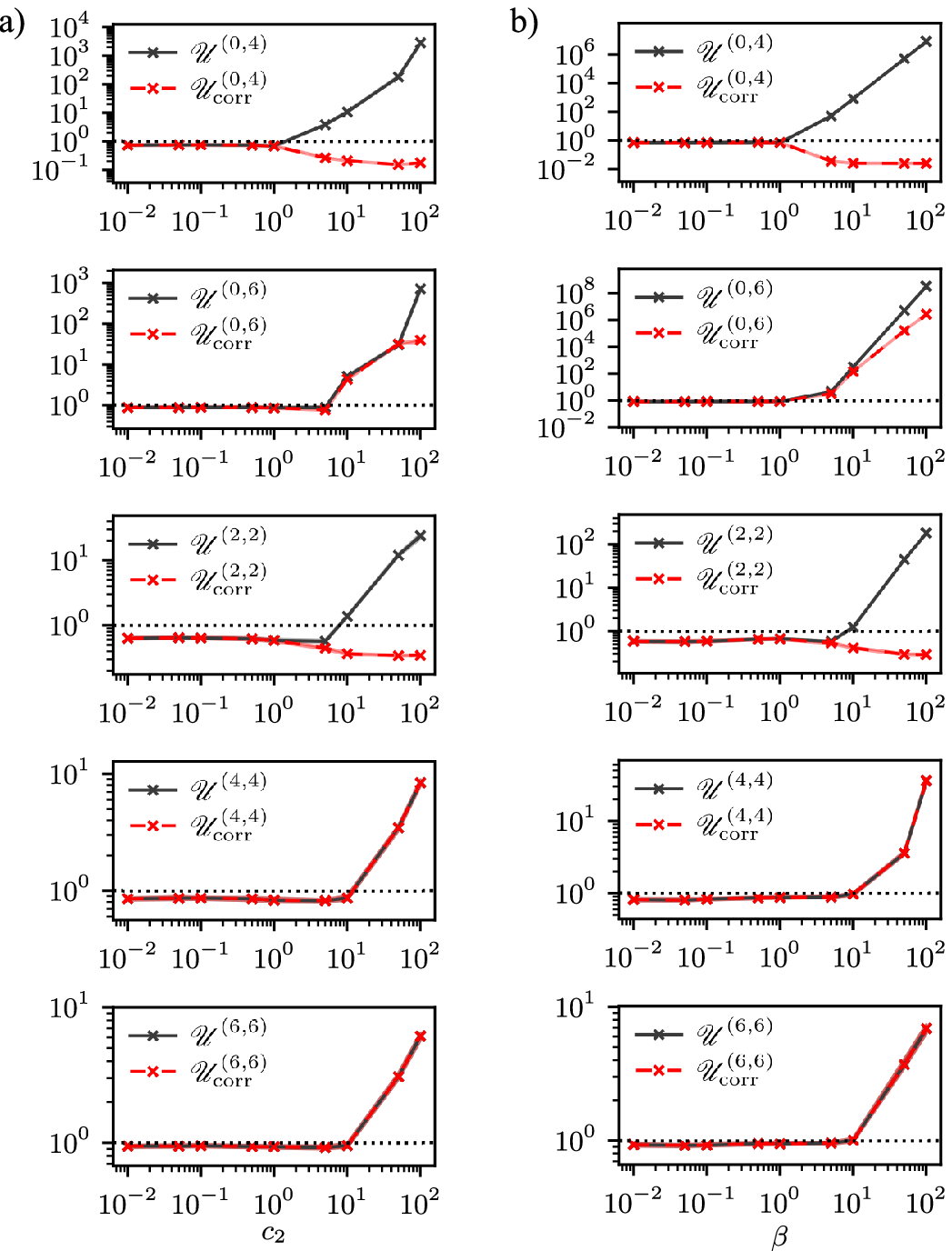}
\caption{
Reconstruction accuracies with ($\umbrae_\text{corr}^{(\ell,m)}$; red) and without ($\umbrae^{(\ell,m)}$; black) integrating correction terms of order $\mathcal{O}(\dt^2)$ for various conditional moments (cf.~\fref{fig:example_coupling}a and~\fref{fig:results_umbrae}).
a) Bivariate jump-diffusion model with uni-directional couplings (see equation~\eref{eq:2djdcoupled}) for various values of coupling strength $c_2$.
b) Bivariate jump-diffusion model (see equation~\eref{eq:2djdmixing}) for various values of the diffusion-scaling parameter $\scalingb$.
The horizontal dotted line indicates a sufficient accuracy. 
Medians and interquartile ranges (shaded area) derived from 50 time series generated with the respective models using random initial conditions.
Lines are for eye guidance only.
}
\label{fig:umbrae_comparison}
\end{figure}
For the data shown in \fref{fig:example_corrected}a, the correction terms of order $\mathcal{O}(\dt^2)$ can be of the same or even greater magnitude than terms of order~$\mathcal{O}(\dt)$ and thus have a non-negligible effect on the accuracy of the reconstruction of conditional moments.
Particularly the accuracy of the reconstruction of moments $\Ktwo{0}{4}$ and $\Ktwo{2}{2}$ is considerably improved ($\umbrae_\text{corr}^{(0,4)}<1$ and $\umbrae_\text{corr}^{(2,2)}<1$; see \fref{fig:umbrae_comparison}) even at large values of the coupling strength or at large values of the diffusion-scaling parameter.
However, we still observe inaccuracies in the reconstruction of conditional moments $\Ktwo{0}{6}$, $\Ktwo{4}{4}$, and $\Ktwo{6}{6}$, and we expect that considering terms of higher order of $\dt$ ($\mathcal{O}(\dt^i)$, $i\geq3$) will further improve the accuracy of the reconstruction of these conditional moments.

%%%%%%%%%%%%%%%%%%%%%%%%%%%%%%%%%%%%%%%%%%%%%%%%%%%%%%%%%%%%%%%%%%%%%%%%
\section{Concluding remarks}
\label{sec:conclusions}
%%%%%%%%%%%%%%%%%%%%%%%%%%%%%%%%%%%%%%%%%%%%%%%%%%%%%%%%%%%%%%%%%%%%%%%%

We evaluate the significance of a bivariate jump-diffusion model for a data-driven characterization of interactions between complex dynamical systems.
Investigating various coupled and non-coupled jump-diffusion processes, we observed strong deviations between 
conditional moments of the underlying jump-diffusion model and those estimated from time-series data  
and conjectured that these deviations result from the finiteness of the sampling interval.
We derived correction terms for conditional moments in higher orders of the sampling interval and could demonstrate that these corrections strongly enhance the accuracy of a data-driven reconstruction of stochastic evolution equations from time-series data in terms of bivariate jump-diffusion models.

Our findings demonstrate that the drift, diffusion and jumps induce terms of order $\mathcal{O}(\dt^2)$ in all conditional moments and are most pronounced in conditional moments with jump contributions (orders $\geq4$).
A blending of all parts of the dynamics should thus be taken into account when investigating interacting jump-diffusion processes.
To further enhance the significance of the bivariate jump-diffusion model for the analysis of empirical data, future studies should investigate other possible influencing factors such as measurement noise, limited observation time (finite number of data points), or the impact of indirect interactions mediated by observed/unobserved additional processes. 

\ack
We are grateful to M. Reza Rahimi Tabar for constructive discussions and valuable comments.

%%%%%%%%%%%%%%%%%%%%%%%%%%%%%%%%%%%%%%%%%%%%%%%%%%%%%%%%%%%%%%%%%%%%%%%%

\clearpage

%%%%%%%%%%%%%%%%%%%%%%%%%%%%%%%%%%%%%%%%%%%%%%%%%%%%%%%%%%%%%%%%%%%%%%%%
\section*{Appendix}
\appendix
\setcounter{section}{0}
%%%%%%%%%%%%%%%%%%%%%%%%%%%%%%%%%%%%%%%%%%%%%%%%%%%%%%%%%%%%%%%%%%%%%%%%

%%%%%%%%%%%%%%%%%%%%%%%%%%%%%%%%%%%%%%%%%%%%%%%%%%%%%%%%%%%%%%%%%%%%%%%%
\section{Scale-independent measure to assess the accuracy of a data-driven reconstruction of conditional moments}
\label{app:measures}
%%%%%%%%%%%%%%%%%%%%%%%%%%%%%%%%%%%%%%%%%%%%%%%%%%%%%%%%%%%%%%%%%%%%%%%%

For each two-dimensional conditional moments of order $(\ell, m)$, we consider the unscaled mean bounded relative absolute error~\cite{chen2017}
\begin{eqnarray*}
\umbrae^{(\ell,m)} =\frac{\boundedrelativeerror^{(\ell,m)} }{1-\boundedrelativeerror^{(\ell,m)} },
\label{eq:umbrae}
\end{eqnarray*}
with the weighted average of bounded relative errors 
\begin{eqnarray*}
\boundedrelativeerror^{(\ell,m)} =\sum\limits_{i,j=1}^B p(\obs_{1,i},\obs_{2,j})\frac{\Big|\Delta_{ij}^{(\ell,m)} \Big|}{\Big|\Delta_{ij}^{(\ell,m)} \Big|+\Big|\Ktwo{\ell}{m}(\obs_{1,i},\obs_{2,j},\dt)\Big|},
\end{eqnarray*}
where $\Delta_{ij}^{(\ell,m)}=\estKtwo{\ell}{m}(\obs_{1,i},\obs_{2,j},\dt) - \Ktwo{\ell}{m}(\obs_{1,i},\obs_{2,j},\dt)$ is the difference between the estimated and theoretical conditional moment,
$p(\cdot)$ the estimated probability density that is used as normalized weight, and $B$ the number of bins for each dimension.

$\umbrae^{(\ell,m)}< 1$ indicates a sufficient accuracy of the reconstruction of conditional moments of order $(\ell,m)$ in the sense that $\Delta_{ij}^{(\ell,m)}$ is smaller than $\Ktwo{\ell}{m}(\obs_{1,i},\obs_{2,j},\dt)$ on average.
We note that the value of $\umbrae^{(\ell,m)}$ becomes large or undefined if the value of the theoretical conditional moments tends to $0$, which can lead to a misinterpretation. 

In our histogram-based investigations, we used $B=20$ bins for each dimension and considered a range of $\pm \sigma_i$ for each $\obs_i$. 
We refer to Ref.~\cite{knuth2019} for a discussion on the optimal choice of the number of bins and to Refs.~\cite{lamouroux2009,gorjao-meirinhos2019} for other, e.g., kernel-based estimation techniques.

%%%%%%%%%%%%%%%%%%%%%%%%%%%%%%%%%%%%%%%%%%%%%%%%%%%%%%%%%%%%%%%%%%%%%%%%
\section{Derivation of conditional moments of bivariate jump-diffusion models for different orders of $\dt$}
\label{app:timeintervalderivation-jumps}
%%%%%%%%%%%%%%%%%%%%%%%%%%%%%%%%%%%%%%%%%%%%%%%%%%%%%%%%%%%%%%%%%%%%%%%%

We follow Refs.~\cite{lehnertz2018,tabar2019book} to derive conditional moments of bivariate jump-diffusion models for different orders of $\dt$ using the Kramers-Moyal adjoint operator.
With the abbreviations
\begin{eqnarray*}
\atwo{1}{0} &= h_1\\
\atwo{0}{1} &= h_2\\
\btwo{1}{1} &= \frac{1}{2}\Big[\diffusion_{11}\diffusion_{21}+\diffusion_{12}\diffusion_{22}\Big] \\
\btwo{2}{0} &= \frac{1}{2}\Big[\diffusion_{11}^2+\jumpsize_{11}\jumprate_1+\diffusion_{12}^2+\jumpsize_{12}\jumprate_2\Big] \\
\btwo{0}{2} &= \frac{1}{2}\Big[\diffusion_{21}^2+\jumpsize_{21}\jumprate_1+\diffusion_{22}^2+\jumpsize_{22}\jumprate_2\Big] \\
\ctwo{2\ell}{2m} &= \frac{1}{(2\ell+2m)!}\Big[\jumpsize_{11}^{\ell}\jumpsize_{21}^{m}\jumprate_1+\jumpsize_{12}^{\ell}\jumpsize_{22}^{m}\jumprate_2\Big]\frac{(2\ell)!}{2^{\ell} \ell!}\frac{(2m)!}{2^{m}m!}, \\
\end{eqnarray*}
where $(\ell,m) \in \mathbb{N}^+$, one can find the following corrections for conditional moments of orders $\ell=m=6$ of a bivariate jump-diffusion model:
\begin{eqnarray*}
\fl\Ktwo{1}{0}(\dobs,t,\dt) = & \atwo{1}{0}\dt\\
& + \frac{1}{2}\Big[ \atwo{1}{0}\del{1}\atwo{1}{0} + \atwo{0}{1}\del{2}\atwo{1}{0}\\
& + \btwo{2}{0}\del{1}^2\atwo{1}{0} + \btwo{0}{2}\del{2}^2\atwo{1}{0} + 2\btwo{1}{1}\del{1}\del{2}\atwo{1}{0}\\
& + 6\ctwo{2}{2}\del{1}^2\del{2}^2\atwo{1}{0}
+ \mathcal{O}(\delta)\Big]\dt^2 + \mathcal{O}(\dt^3)\\
\fl\Ktwo{2}{0}(\dobs,t,\dt) =& 2\btwo{2}{0}\dt\\
& + \frac{1}{2}\Big[ 2\kl{\atwo{1}{0}}^2 + 2\kl{\atwo{1}{0}\del{1}\btwo{2}{0} + \atwo{0}{1}\del{2}\btwo{2}{0}}\\
& + 4\kl{\btwo{2}{0}\del{1}\atwo{1}{0} + \btwo{1}{1}\del{2}\atwo{1}{0}}\\
& + 2\kl{\btwo{2}{0}\del{1}^2\btwo{2}{0} + \btwo{0}{2}\del{2}^2\btwo{2}{0} + 2\btwo{1}{1}\del{1}\del{2}\btwo{2}{0}}\\
& + 8 \big(\ctwo{4}{0}\del{1}^3\atwo{1}{0} + 3\ctwo{2}{2}\del{1}\del{2}^2\atwo{1}{0}\big)\\
& + 12\ctwo{2}{2}\del{1}^2\del{2}^2\btwo{2}{0}
+ 5!\ctwo{4}{2}\del{1}^3\del{2}^2\atwo{1}{0}\\
& + \mathcal{O}(\delta)\Big]\dt^2 + \mathcal{O}(\dt^3)\\
\fl\Ktwo{4}{0}(\dobs,t,\dt) = & 4!\ctwo{4}{0}\dt\\
& + \frac{1}{2}\Big[  4!\big(\btwo{2}{0}\big)^2
+ 4!\big(\atwo{1}{0}\del{1}\ctwo{4}{0} + \atwo{0}{1}\del{2}\ctwo{4}{0}\big)\\
& + 4\cdot4!\ctwo{4}{0}\del{1}\atwo{1}{0}
+ 4!\big( \btwo{2}{0}\del{1}^2\ctwo{4}{0} + \btwo{0}{2}\del{2}^2\ctwo{4}{0}\\
& + 2\btwo{1}{1}\del{1}\del{2}\ctwo{4}{0} \big)
+ 6\cdot4!\big( \ctwo{4}{0}\del{1}^2\btwo{2}{0} + \ctwo{2}{2}\del{2}^2\btwo{2}{0} \big)\\
& + 4\cdot5! \big( \ctwo{6}{0}\del{1}^3\atwo{1}{0} + 3\ctwo{4}{2}\del{1}\del{2}^2\atwo{1}{0} \big)\\
& + 6\cdot4!\ctwo{2}{2}\del{1}^2\del{2}^2\ctwo{4}{0}
 + 3\cdot6!\ctwo{4}{2}\del{1}^2\del{2}^2\btwo{2}{0}
+ \mathcal{O}(\delta)\Big]\dt^2 + \mathcal{O}(\dt^3)\\
\fl\Ktwo{6}{0}(\dobs,t,\dt) = & 6!\ctwo{6}{0}\dt\\
& + \frac{1}{2}\Big[ 2\cdot6!\btwo{2}{0}\ctwo{4}{0} + 6!\big( \atwo{1}{0}\del{1}\ctwo{6}{0} + \atwo{0}{1}\del{2}\ctwo{6}{0} \big)\\
& + 6\cdot6!\ctwo{6}{0}\del{1}\atwo{1}{0} + 6!\big( \btwo{2}{0}\del{1}^2\ctwo{6}{0} + \btwo{0}{2}\del{2}^2\ctwo{6}{0}\\
& + 2\btwo{1}{1}\del{1}\del{2}\ctwo{6}{0} \big)
+ 6\cdot6!\big( \ctwo{4}{0}\del{1}^2\ctwo{4}{0} + \ctwo{2}{2}\del{2}^2\ctwo{4}{0} \big)\\
& + \frac{(6!)^2}{2!4!}\big( \ctwo{6}{0}\del{1}^2\btwo{2}{0} + \ctwo{4}{2}\del{2}^2\btwo{2}{0} \big)\\
& + 6\cdot\frac{8!}{3!}\big( \ctwo{8}{0}\del{1}^3\atwo{1}{0} + 3\ctwo{6}{2}\del{1}\del{2}^2\atwo{1}{0} \big)\\
& + 6\cdot6!\ctwo{2}{2}\del{1}^2\del{2}^2\ctwo{6}{0}
+ 6\cdot\frac{(6!)^2}{4!2!}\ctwo{4}{2}\del{1}^2\del{2}^2\ctwo{4}{0}\\
& + 30\cdot\frac{8!}{2!2!}\ctwo{6}{2}\del{1}^2\del{2}^2\btwo{2}{0}
+ \mathcal{O}(\delta)\Big]\dt^2 + \mathcal{O}(\dt^3)
\end{eqnarray*}
\begin{eqnarray*}
\fl\Ktwo{1}{1}(\dobs,t,\dt) =& 2\btwo{1}{1}\dt\\
& + \frac{1}{2}\Big[ 2\atwo{1}{0}\atwo{0}{1} + 2\big(\atwo{1}{0}\del{1}\btwo{1}{1} + \atwo{0}{1}\del{2}\btwo{1}{1}\big)\\
& + 2\big(\btwo{2}{0}\del{1}\atwo{0}{1} + \btwo{1}{1}\del{2}\atwo{0}{1} + \btwo{1}{1}\del{1}\atwo{1}{0} + \btwo{0}{2}\del{2}\atwo{1}{0} \big)\\
& + 2\big(\btwo{2}{0}\del{1}^2\btwo{1}{1} + \btwo{0}{2}\del{2}^2\btwo{1}{1} + 2\btwo{1}{1}\del{1}\del{2}\btwo{1}{1}\big)\\
& + 4\big(\ctwo{4}{0}\del{1}^3\atwo{0}{1} + 3\ctwo{2}{2}\del{1}\del{2}^2\atwo{0}{1} +  \ctwo{0}{4}\del{2}^3\atwo{1}{0}\\
& + 3\ctwo{2}{2}\del{1}^2\del{2}\atwo{1}{0}\big) + 12\ctwo{2}{2}\del{1}^2\del{2}^2\btwo{1}{1}\\
& + \frac{5!}{2!}\big( \ctwo{4}{2}\del{1}^3\del{2}^2\atwo{0}{1} + \ctwo{2}{4}\del{1}^2\del{2}^3\atwo{1}{0}\big)
+ \mathcal{O}(\delta)\Big]\dt^2 + \mathcal{O}(\dt^3)\\
\fl\Ktwo{2}{2}(\dobs,t,\dt) = & 4!\ctwo{2}{2}\dt\\
& + \frac{1}{2}\Big[ \frac{4!}{3}\big(\btwo{2}{0}\btwo{0}{2} + 2\big(\btwo{1}{1}\big)^2\big)\\
& + 4!\big(\atwo{1}{0}\del{1}\ctwo{2}{2} + \atwo{0}{1}\del{2}\ctwo{2}{2}\big)\\
& + 2\cdot4!\big( \ctwo{2}{2}\del{1}\atwo{1}{0} + \ctwo{2}{2}\del{2}\atwo{0}{1} \big)\\
& + 4!\big( \btwo{2}{0}\del{1}^2\ctwo{2}{2} + \btwo{0}{2}\del{2}^2\ctwo{2}{2} + 2\btwo{1}{1}\del{1}\del{2}\ctwo{2}{2} \big)\\
& + 4!\big( \ctwo{4}{0}\del{1}^2\btwo{0}{2} + \ctwo{2}{2}\del{2}^2\btwo{0}{2} + \ctwo{2}{2}\del{1}^2\btwo{2}{0}  + \ctwo{0}{4}\del{2}^2\btwo{2}{0}\\
& + 8\ctwo{2}{2}\del{1}\del{2}\btwo{1}{1} \big)
+ 2\cdot\frac{6!}{3!}\big( \ctwo{4}{2}\del{1}^3\atwo{1}{0} + 3\ctwo{2}{4}\del{1}\del{2}^2\atwo{1}{0}\\
& + \ctwo{2}{4}\del{2}^3\atwo{0}{1}
+ 3\ctwo{4}{2}\del{1}^2\del{2}\atwo{0}{1} \big)
+ 6\cdot4!\ctwo{2}{2}\del{1}^2\del{2}^2\ctwo{2}{2}\\
& + \frac{5!}{2!}\big( 6\big(\ctwo{4}{2}\del{1}^2\del{2}^2\btwo{0}{2} + \ctwo{2}{4}\del{1}^2\del{2}^2\btwo{2}{0}\big) \\
& + 16\big(\ctwo{4}{2}\del{1}^3\del{2}\btwo{1}{1} + \ctwo{2}{4}\del{1}\del{2}^3\btwo{1}{1}\big) \big)
+ \mathcal{O}(\delta)\Big]\dt^2 + \mathcal{O}(\dt^3)\\
\fl\Ktwo{4}{4}(\dobs,t,\dt) = & 8!\ctwo{4}{4}\dt\\
& + \frac{1}{2}\Big[ 12\cdot6!\big(\btwo{2}{0}\ctwo{2}{4} + \btwo{0}{2}\ctwo{4}{2}\big)\\
& + 2\cdot4!4!\big(\ctwo{4}{0}\ctwo{0}{4} + 18\big(\ctwo{2}{2}\big)^2\big)\\
& + 12\cdot6!\big(\ctwo{2}{4}\btwo{2}{0} + \ctwo{4}{2}\btwo{0}{2} \big) 
+ 8!\big( \atwo{1}{0}\del{1}\ctwo{4}{4} + \atwo{0}{1}\del{2}\ctwo{4}{4} \big)\\
& + 4\cdot8!\big(\ctwo{4}{4}\del{1}\atwo{1}{0} + \ctwo{4}{4}\del{2}\atwo{0}{1}\big)\\
& + 8!\big( \btwo{2}{0}\del{1}^2\ctwo{4}{4} + \btwo{0}{2}\del{2}^2\ctwo{4}{4} + 2\btwo{1}{1}\del{1}\del{2}\ctwo{4}{4} \big)\\
& + 3\cdot4!6!\big( \ctwo{4}{0}\del{1}^2\ctwo{2}{4} + \ctwo{2}{2}\del{2}^2\ctwo{2}{4} + \ctwo{0}{4}\del{2}^2\ctwo{4}{2}\\
& + \ctwo{2}{2}\del{1}^2\ctwo{4}{2} \big)\\
& + 12\cdot6!\big(\ctwo{6}{0}\del{1}^2\ctwo{0}{4} + \ctwo{4}{2}\del{2}^2\ctwo{0}{4}
+ \ctwo{2}{4}\del{1}^2\ctwo{4}{0} + \ctwo{0}{6}\del{2}^2\ctwo{4}{0}\\
& + 36\big(\ctwo{4}{2}\del{1}^2\ctwo{2}{2} + \ctwo{2}{4}\del{2}^2\ctwo{2}{2}\big)\big)\\
& + 6\cdot8!\big(\ctwo{4}{4}\del{1}^2\btwo{2}{0} + \ctwo{2}{6}\del{2}^2\btwo{2}{0} + \ctwo{6}{2}\del{1}^2\btwo{0}{2}  + \ctwo{4}{4}\del{2}^2\btwo{0}{2}\\
& + 16\cdot\frac{2!}{3!}\ctwo{4}{4}\del{1}\del{2}\btwo{1}{1}\big)
+ 6\cdot8!\ctwo{2}{2}\del{1}^2\del{2}^2\ctwo{4}{4}\\
& + 6\cdot8! \big( \ctwo{6}{2}\del{1}^2\del{2}^2\ctwo{0}{4} + \ctwo{2}{6}\del{1}^2\del{2}^2\ctwo{4}{0}
+ 36\ctwo{4}{4}\del{1}^2\del{2}^2\ctwo{2}{2}\big)\\
& + \mathcal{O}(\delta)\Big]\dt^2 + \mathcal{O}(\dt^3)
\end{eqnarray*}
\begin{eqnarray*}
\fl\Ktwo{6}{6}(\dobs,t,\dt) = & 12!\ctwo{6}{6}\dt\\
& + \frac{1}{2}\Big[ 30\cdot10!\big(\btwo{2}{0}\ctwo{4}{6} + \btwo{0}{2}\ctwo{6}{4}\big)\\
& + \frac{6!8!}{2!}\big( \ctwo{4}{0}\ctwo{2}{6} + \ctwo{0}{4}\ctwo{6}{2} + 15\ctwo{2}{2}\ctwo{4}{4} \big)\\
& + 2\cdot6!6!\big( \ctwo{6}{0}\ctwo{0}{6} + \frac{(6!)^2}{(2!4!)^2}\ctwo{2}{4}\ctwo{4}{2} \big)\\
& + \frac{6!8!}{2!}\big( \ctwo{2}{6}\ctwo{4}{0} + \ctwo{6}{2}\ctwo{0}{4} + 15\ctwo{4}{4}\ctwo{2}{2} \big)\\
& + 30\cdot10!\big( \ctwo{4}{6}\btwo{2}{0} + \ctwo{6}{4}\btwo{0}{2} \big)\\
& + 12!\big(\atwo{1}{0}\del{1}\ctwo{6}{6} + \atwo{0}{1}\del{2}\ctwo{6}{6}\big)
+ 6\cdot12!\big(\ctwo{6}{6}\del{1}\atwo{1}{0} + \ctwo{6}{6}\del{2}\atwo{0}{1}\big)\\
& + 12!\big(\btwo{2}{0}\del{1}^2\ctwo{6}{6} + \btwo{0}{2}\del{2}^2\ctwo{6}{6} + 2\btwo{1}{1}\del{1}\del{2}\ctwo{6}{6}\big)\\
& + \frac{6!10!}{2!2!}\big( \ctwo{4}{0}\del{1}^2\ctwo{4}{6} + \ctwo{2}{2}\del{2}^2\ctwo{4}{6} + \ctwo{2}{2}\del{1}^2\ctwo{6}{4} + \ctwo{0}{4}\del{2}^2\ctwo{6}{4} \big)\\
& + 30\cdot\frac{6!8!}{2!}\big( \ctwo{6}{0}\del{1}^2\ctwo{2}{6} + \ctwo{4}{2}\del{2}^2\ctwo{2}{6} + \ctwo{2}{4}\del{1}^2\ctwo{6}{2} + \ctwo{0}{6}\del{2}^2\ctwo{6}{2}\\
& + 15\big( \ctwo{4}{2}\del{1}^2\ctwo{4}{4} + \ctwo{2}{4}\del{2}^2\ctwo{4}{4} \big) \big)\\
& + \frac{6!8!}{2!}\Big(\ctwo{8}{0}\del{1}^2\ctwo{0}{6} + \ctwo{6}{2}\del{2}^2\ctwo{0}{6} + \ctwo{2}{6}\del{1}^2\ctwo{6}{0} + \ctwo{0}{8}\del{2}^2\ctwo{6}{0}\\
& + \frac{(6!)^2}{(4!2!)^2}\big(\ctwo{4}{4}\del{1}^2\ctwo{4}{2} + \ctwo{2}{6}\del{2}^2\ctwo{4}{2}+\ctwo{6}{2}\del{1}^2\ctwo{2}{4} + \ctwo{4}{4}\del{2}^2\ctwo{2}{4}\big)\Big)\\
& + \frac{6!10!}{2!2!}\big(\ctwo{4}{6}\del{1}^2\ctwo{4}{0} + \ctwo{2}{8}\del{2}^2\ctwo{4}{0} + \ctwo{8}{2}\del{1}^2\ctwo{0}{4} + \ctwo{6}{4}\del{2}^2\ctwo{0}{4}\\
& + 15\big(\ctwo{6}{4}\del{1}^2\ctwo{2}{2}+\ctwo{4}{6}\del{2}^2\ctwo{2}{2}\big)\big)\\
& + 15\cdot12!\big(\ctwo{8}{4}\del{1}^2\btwo{0}{2} + \ctwo{6}{6}\del{2}^2\btwo{0}{2} + \ctwo{6}{6}\del{1}^2\btwo{2}{0} + \ctwo{4}{8}\del{2}^2\btwo{2}{0}\\
& + \frac{4!4!}{5!}\ctwo{6}{6}\del{1}\del{2}\btwo{1}{1}\big)\\
& + 6\cdot12!\ctwo{2}{2}\del{1}^2\del{2}^2\ctwo{6}{6}\\
& + 30\cdot7!8!\big(\ctwo{6}{2}\del{1}^2\del{2}^2\ctwo{2}{6} + \ctwo{2}{6}\del{1}^2\del{2}^2\ctwo{6}{2} + 15\ctwo{4}{4}\del{1}^2\del{2}^2\ctwo{4}{4}\big)\\
& + \frac{(6!)^2}{(4!2)^2}\frac{6!10!}{2!2!}\big(\ctwo{4}{6}\del{1}^2\del{2}^2\ctwo{4}{2} + \ctwo{6}{4}\del{1}^2\del{2}^2\ctwo{2}{4}\big)\\
& + 90\cdot12!\big( \ctwo{4}{8}\del{1}^2\del{2}^2\ctwo{4}{0} +  \ctwo{8}{4}\del{1}^2\del{2}^2\ctwo{0}{4} + 15\ctwo{6}{6}\del{1}^2\del{2}^2\ctwo{2}{2}\big)\\
& + \mathcal{O}(\delta)\Big]\dt^2 + \mathcal{O}(\dt^3)
\end{eqnarray*}
For the differential operator, we use the short notation $\del{i}=\frac{\partial}{\partial\obs_i}$.
With $\mathcal{O}(\delta)$, we indicate all terms that contain $\ctwo{\ell}{m}$ of higher-order or derivatives $\del{i}^j$, $j>3$,
and with $\mathcal{O}(\dt^3)$ all terms that contain higher orders $(\ge3)$ of the sampling interval $\dt$.
One obtains $\Ktwo{m}{\ell}(\dobs,t,\dt)$ from $\Ktwo{\ell}{m}(\dobs,t,\dt)$ by interchanging the indices $i\in\{1,2\}$ of the differential operator $\del{i}\equiv\frac{\partial}{\partial_{\obs_i}}$ and by interchanging $\ell$ with $m$ in the orders of $\atwo{\ell}{m}$, $\btwo{\ell}{m}$, and $\ctwo{\ell}{m}$.

%%%%%%%%%%%%%%%%%%%%%%%%%%%%%%%%%%%%%%%%%%%%%%%%%%%%%%%%%%%%%%%%%%%%%%%%
\section*{References}
%%%%%%%%%%%%%%%%%%%%%%%%%%%%%%%%%%%%%%%%%%%%%%%%%%%%%%%%%%%%%%%%%%%%%%%%
\providecommand{\newblock}{}

\end{document}